\documentclass[fleqn,preprint,showpacs,preprintnumbers,amsmath,amssymb]{revtex4}
\usepackage{graphicx}
\usepackage{dcolumn}
\usepackage{bm}

\begin{document}

\title{Heavy-Fermion Instability in Double-Degenerate Plasmas}
\author{M. Akbari-Moghanjoughi}
\affiliation{Azarbaijan University of
Tarbiat Moallem, Faculty of Sciences, Department of Physics,
51745-406, Tabriz, Iran}

\date{\today}
\begin{abstract}
In this work we study the propagations of normal frequency modes for quantum hydrodynamic (QHD) waves in the linear limit and introduce a new kind of instability in a double-degenerate plasma. Three different regimes, namely, low, intermediate and high magnetic field strengths are considered which span the applicability of the work to a wide variety of environments. Distinct behavior is observed for different regimes, for instance, in the laboratory-scale field regime no frequency-mode instability occurs unlike those of intermediate and high magnetic-field strength regimes. It is also found that the instability of this kind is due to the heavy-fermions which appear below a critical effective-mass parameter ($\mu_{cr}=\sqrt{3}$) and that the responses of the two (lower and upper frequency) modes to fractional effective-mass change in different effective-mass parameter ranges (below and above the critical value) are quite opposite to each other. It is shown that, the heavy-fermion instability due to extremely high magnetic field such as that encountered for a neutron-star crust can lead to confinement of stable propagations in both lower and upper frequency modes to the magnetic poles. Current study can have important implications for linear wave dynamics in both laboratory and astrophysical environments possessing high magnetic fields.
\end{abstract}

\maketitle

\section{Introduction}

Since the early developments \cite{bohm, pines, levine}, more investigations have appeared on the problems of wave dynamics in quantum plasmas \cite{bonitz, gardner, haas1, Markowich, manfredi, shukla, akbari1}, a degenerate dense and cold ionized matter. A quantum plasma is characterized by the state of plasma in which the thermal energy of the constituent particles fall below a well-defined energy called the Fermi-energy and the particle waves overlap causing a new type collective phenomenon through the Pauli exclusion rule \cite{kittel}. It has been shown that, in a quantum plasma the nonlinear response to plasma excitations may differ fundamentally due to effects such as electron nonlocality \cite{haas2}, degeneracy \cite{akbari2}, and spin-orbit magnetization \cite{marklund}. The nonlocality effects due to the Bohm-force has been shown to lead to some criticality in the quantum hydrodynamics (QHD) \cite{haas3, akbari3}. It has also been revealed that the magnetohydrodynamics MHD waves substantially altered when the normal degeneracy is changed to the Chandrasekhar relativistic degeneracy \cite{akbari4} in a Thomas-Fermi model and different degeneracy-regimes in a quantum plasma can lead to different QHD-wave behavior when a strong magnetic field is involved \cite{akbari5}. On the other hand, the newly developed extended MHD \cite{marklund1, brodin, markshukla} confirms that the plasma spin-polarization can lead to even more peculiar features in a quantum plasma. It has been shown that, the plasma magnetization which introduces a negative like pressure term can have significant effect on nonlinear density excitation in quantum plasmas \cite{marklund2} and even alter the Jeans instability \cite{lund}. More recently a fully relativistic hydrodynamic model including the spin effect has been proposed by F. A. Asenjo et al. \cite{asenjo} based on particle-antiparticle Dirac theory.

In the presence of strong magnetic field the quantum plasma can undergo a double quantum-pressure mechanisms \cite{can}, called the double-degeneracy analogous to double-adiabatic \cite{krall} process, due to the electron spin and orbital quantization. It has be shown that such stress-anisotropy can lead to a transverse plasma magnetic collapse in strongly magnetized plasmas \cite{chai}. Because of the transverse electron momentum quantization under the magnetic-field in addition to normal electron degeneracy new energy binding mechanisms may occur in a quantum plasma \cite{dong}. Two distinct magnetization mechanisms interplay delicately in a magnetization process, namely, the positive Pauli paramagnetism and the Landau negative diamagnetism \cite{landau}. It has been shown that, the dominance of each of these mechanisms can lead to quite different nonlinear behavior in a quantum plasma \cite{akbari6}. According to the value of the electron effective mass there are regimes in which the plasma becomes paramagnetic or diamagnetic \cite{sho}. A vary large value of effective mass-ration ($m_e/m_e^*$) has been recently observed in Bismuth compounds \cite{zhang}. On the other hand, for the superconductor heavy-fermion systems this value can be very small giving rise to the quantum criticality \cite{nakat}. In this investigation we show that such heavy-fermion quantum-plasma systems can develop a new kind of instability in MHD wave propagations. We consider different regimes of magnetic-field/density to show the relevance of our investigation to wide environments from laboratory to the astrophysical dense plasmas. The presentation of the article is as follows. The QHD plasma model including the spin contribution is introduced in Sec. \ref{equations}. The linear dispersion of spin-induced magnetoacoustic waves are derived in Sec. \ref{extreme} and the dispersion and instabilities are discussed based on field/density regimes. Finally, a conclusions are drawn in Sec. \ref{con}.

\section{Double-Degenerate QHD Model}\label{equations}

\textbf{The full quantum magnetohydrodynamic fluid set of equations consists of continuity, momentum, Poisson's and Maxwell electromagnetic equations which has been derived in Ref. \cite{haas4} for a general collisional quantum plasma. In the following we focus on the simplified model of uniform (in space and time) magnetization which eliminates the role of electromagnetic interactions. Therefore, our simplified two-fluid quasineutral plasma model consists of contributions from the generalized pressure tensor, ${\bf{\tilde P}}$, nonlocality (quantum diffraction) effect called the Bohma-force, $\bf{F_Q}$, and the Lorentz force, $\bf{F_L}$. We further assume a collision-less condition for the plasma due to the Pauli blocking mechanism which greatly reduces the fermion collision frequency in a quantum plasma compared to the ordinary ones. For a singly-ionized collisionless quantum plasma the quantum hydrodynamic fluid equation governing the electrostatic perturbations, reads as}
\begin{equation}\label{dimensional0}
\begin{array}{l}
\frac{{\partial {n_{i,e}}}}{{\partial t}} + \nabla \cdot({n_{i,e}}{{\bf{u}}_{{\bf{i}},{\bf{e}}}}) = 0,\hspace{3mm}{{\bf{u}}_{({\bf{i}},{\bf{e}})}} \equiv [{u_{x(i,e)}},{u_{y(i,e)}},{u_{z(i,e)}}], \\
{m_{i,e}}\frac{{d{{\bf{u}}_{{\bf{i,e}}}}}}{{dt}} =  \mp e({\bf{E}} + {{\bf{u}}_{{\bf{i,e}}}} \times {{\bf{B}}_{\bf{0}}}) - \frac{1}{{{n_{i,e}}}}\nabla \cdot{{{\bf{\tilde P}}}_{i,e}}({{\rm{n}}_{i,e}}) + \frac{{{\hbar ^2}}}{{2{m_{i,e}}}}\nabla \frac{{\Delta \sqrt {{n_{i,e}}} }}{{\sqrt {{n_{i,e}}} }}, \\
\nabla \cdot{\bf{E}} = 4\pi{\rho }. \\
\end{array}
\end{equation}
\textbf{Considering the collisionless quasineutral uniformly magnetized quantum plasma consisting of degenerate-electrons and classical nondegenerate-ions, we further simplify the model by neglecting the pressure due to ions compared to that of electron degeneracy pressure. The arbitrary strength uniform magnetic field, ${\bf{B_0}}$, is assumed to be directed along the $x$ axis. Therefore, the complete set of simplified QHD equations describing the dynamics of electrons and ions in magnetized degenerate plasma, neglecting the Bohm-force on classical ions (due to the much lower momentum of ions compared to that of electrons), can be written as}
\begin{equation}\label{dimensional}
\begin{array}{l}
\frac{{\partial {n_{i,e}}}}{{\partial t}} + \nabla \cdot({n_{i,e}}{{\bf{u}}_{{\bf{i}},{\bf{e}}}}) = 0,\\
\frac{{d{{\bf{u}}_{\bf{i}}}}}{{dt}} =  - \frac{e}{{{m_i}}}\nabla \phi  + {\omega _{ci}}({{\bf{u}}_{\bf{i}}} \times {{\bf{\hat x}}}), \\
\frac{{{m_e}}}{{{m_i}}}\left[ {\frac{{d{{\bf{u}}_e}}}{{dt}} + {\omega _{ce}}({{\bf{u}}_e} \times {{\bf{\hat x}}})} \right] = \frac{e}{{{m_i}}}\nabla \phi  - \frac{1}{{{m_i}{n_e}}}\nabla \cdot{{{\bf{\tilde P}}}_e}({{\rm{n}}_{\rm{e}}}) + \frac{{{\hbar ^2}}}{{2{m_e}{m_i}}}\nabla \frac{{\Delta \sqrt {{n_e}} }}{{\sqrt {{n_e}} }}, \\
\Delta \phi  = 4\pi e({n_e} - {n_i}) \approx 0, \\
\end{array}
\end{equation}
where, $u_{(i,e)}$, $m_{(i,e)}$, $\omega_{c{(i,e)}}=eB_0/m_{(i,e)}c$ and ${{\bf{\tilde P}}_e}({{\rm{n}}_e})$ are the ion/electron velocity, mass, cyclotron frequencies and the electron pressure tensor, respectively. \textbf{In a collisional classical plasma when the characteristic plasma frequencies are lower than the collision frequencies the plasma remains isotropic in all directions. However, in the presence of high magnetic field when the plasma species collisions are rare the plasma isotropy breaks down and the plasma may possess different characteristics parallel and perpendicular to the preferred magnetic field direction. In such a case the pressure tensor can be written as}
\begin{equation}\label{pk}
{\bf{\tilde P}}{\rm{ = }}\left[ {\begin{array}{*{20}{c}}
   {{P_ \bot } + ({P_\parallel } - {P_ \bot }){b_x}{b_x}} & {({P_\parallel } - {P_ \bot }){b_x}{b_y}} & {({P_\parallel } - {P_ \bot }){b_x}{b_z}}  \\
   {({P_\parallel } - {P_ \bot }){b_y}{b_x}} & {{P_ \bot } + ({P_\parallel } - {P_ \bot }){b_y}{b_y}} & {({P_\parallel } - {P_ \bot }){b_y}{b_z}}  \\
   {({P_\parallel } - {P_ \bot }){b_z}{b_x}} & {({P_\parallel } - {P_ \bot }){b_z}{b_y}} & {{P_ \bot } + ({P_\parallel } - {P_ \bot }){b_z}{b_z}}  \\
\end{array}} \right]
\end{equation}
\textbf{with ${\bf{b}}{\rm{ = }}{\bf{B}}{\rm{/}}\left| B \right|$ \cite{krall}. In the presence of a uniform magnetic field the corresponding tensor can be expressed as diagonal with element only parallel and perpendicular to the ambient field as}
\begin{equation}\label{pt}
{{\bf{\tilde P}}}{\rm{ = }}\left[ {\begin{array}{*{20}{c}}
   {{P_{\parallel }}} & 0 & 0  \\
   0 & {{P_{\bot }}} & 0  \\
   0 & 0 & {{P_{\bot }}}  \\
\end{array}} \right],
\end{equation}
\textbf{which introduces a local anisotropy leading to the so-called double-adiabatic Chew-Goldberger-Low (CGL) plasma model \cite{cgl}. Some authors have recently investigated CGL anisotropy effect on the wave dynamics in the classical plasmas \cite{cheong, mahmood}. On the other hand, when the when the plasma species become very close to each other and the particle-waves overlap new collective features emerge due to the spin interactions and a classical plasma may develops quantum hydrodynamic feature \cite{mark}. It has been shown that in the direction transverse to the magnetic field quantum plasma experiences additional negative-like (compared to electron degeneracy force) magnetization force, $\bf{F_m}=B_0\nabla \bf{\Gamma}$, where, $\Gamma$ is the quantum plasma magnetization vector \cite{mark2} which can be considered the total transverse pressure $P_{e\perp}$ as $P_{e\perp}=P_{e\parallel}-\Gamma B_0$ \cite{dong} (e.g. see the footnote in the article). In the reduced quantum plasma model where the motion of electrons are confined to $x$-$y$ plane ($\nabla\equiv [\partial x, \partial y, 0]$), the plasma pressure can be anisotropic perpendicular and parallel to the existing uniform field analogous to the CGL model. However, in this case the anisotropy is not due to real pressure but is due to the plasma magnetization having a purely quantum mechanical origin.}

\textbf{Sondheimer and Wilson \cite{sond} have investigated the diamagnetism due to the electron orbit and paramagnetism due to the electron spin in a zero-temperature ideal Fermi-Dirac gas for an arbitrary magnetic field strength along with oscillatory terms leading to the known de Haas-van Alphen effect \cite{landau}. It is well-known that the plasma magnetization can be due to both electron spin and orbital quantization. }

\textbf{Apart from the oscillatory term the Pauli spin-magnetization per unit volume is given as ${{\Gamma}_P} = \left( {3\mu _B^2{n_e}/2{k_B}{T_{Fe}}} \right){{B_0}}$ (with $T_{Fe}$ being the electron Fermi-temperature related to the electron number density in quantum plasma being much larger than the electron kinetic temperature). On the other hand, a diamagnetic contribution to fully degenerate plasma is caused by the Landau orbital quantization which introduces a new term to plasma magnetization $\Gamma_L=-1/3\Gamma_P$ exists (this expression does not include the electron effective-mass). This leads to the total magnetization of $\Gamma = \Gamma_P + \Gamma_L = n_e\mu _B^2B_0/{E_{Fe}}$ ($E_{Fe}=k_B T_{Fe}=(\hbar^2/2 m_e) (3 \pi^2 n_0)^{2/3}$ is the Fermi-energy) for a ordinary quantum plasma. For many compounds the effective electron mass may differ substantially from that of noninteracting electron mass, leading to a generalized expression for magnetization per unit volume of the ideal zero-temperature quantum plasma, as}
\begin{equation}\label{QP0}
\frac{\Gamma}{V}  = \frac{{{n_e}\mu _B^2}{B_0}}{{{k_B}{T_{Fe}}}}\left[ {1 - \frac{{{\mu ^2}}}{3}} \right],\hspace{3mm}{T_{Fe}} = \frac{{{\hbar ^2}}}{{2{k_B}m_e^*}}{(3{\pi ^2}{n_0})^{2/3}},
\end{equation}
\textbf{with the spin-orbit magnetic susceptibility defined as}
\begin{equation}\label{QP}
{\chi _{so}} = \left[\frac{1}{V}\frac{{\partial \Gamma }}{{\partial B_0}}\right]_{n,V} = \mu _B^2D({E_{Fe}})\left[ {1 - \frac{{{\mu ^2}}}{3}} \right],
\end{equation}
\textbf{where, $\mu={m_e}/{m_e^*}$ is the fractional effective electron-mass parameter with $m_e^{*}$ being the effective electron mass and $D({E_{Fe}}) = 3 n_e /2 E_{Fe}$ the density of electronic states (DoS) at the Fermi level. It is noticed that the susceptibility may be positive or negative depending on the electron effective mass ratio in Eq. (\ref{QP}), hence, the plasma may behave paramagnetic or diamagnetic. It has been shown that, in some Bismuth compounds the fractional effective electron mass can be as high as $\mu\simeq 1000$ \cite{zhang} or some semiconductors. On the other hand, some semiconductors or superconductor heavy-fermion systems may possess the effective-mass ratio values as low as $\mu\simeq 10^{-3}$ \cite{nakat}. In many semiconductors the value of the parameter, $\mu$, contributes to a measurable Landau diamagnetic effects \cite{kittel}. Therefore, defining the Zeeman parameter as, $\epsilon=\mu_B B_0/E_{Fe}$, the magnetization pressure contributes to the below negative transverse force (in $y$-direction) in the normalized momentum equation.}
\begin{equation}\label{so}
{F_{so}} = {B_0}\int {\frac{1}{n}\frac{{\partial \Gamma }}{{\partial n}}} dn =  - \frac{{3{\epsilon^2}}}{2}\left( {1 - \frac{{{\mu ^2}}}{3}} \right)\frac{\partial }{{\partial y}}\ln n.
\end{equation}
\textbf{Now considering the parallel component of the nonrelativistic electron degeneracy pressure as $P_{e\parallel}=(2/5)E_{Fe}n_{0} n_e^{5/3}$ (with $n_0$ being the equilibrium number-density of the electrons), we are led to the normalized and reduced set of QHD equations as below}
\begin{equation}\label{scalar}
\begin{array}{l}
{\partial _t}n + {\partial _x}(n{u_x}) + {\partial _y}(n{u_y}) = 0, \\
{\partial _t}{u_x} + \left( {{u_x}{\partial _x} + {u_y}{\partial _y}} \right){u_x} + {\partial _x}{n^{2/3}} - {H^2}\frac{\partial }{{\partial x}}\left[ {\frac{1}{{\sqrt n }}\left( {\frac{{{\partial ^2}\sqrt n }}{{\partial {x^2}}} + \frac{{{\partial ^2}\sqrt n }}{{\partial {y^2}}}} \right)} \right] = 0, \\
{\partial _t}{u_y} + \left( {{u_x}{\partial _x} + {u_y}{\partial _y}} \right){u_y} + {\partial _y}{n^{2/3}} - \frac{{3{\epsilon^2}}}{2}\left( {1 - \frac{{{\mu ^2}}}{3}} \right)\frac{\partial }{{\partial y}}\ln n \\ - {H^2}\frac{\partial }{{\partial y}}\left[ {\frac{1}{{\sqrt n }}\left( {\frac{{{\partial ^2}\sqrt n }}{{\partial {x^2}}} + \frac{{{\partial ^2}\sqrt n }}{{\partial {y^2}}}} \right)} \right] - \Omega{u_z}= 0, \\
{\partial _t}{u_z} + \left( {{u_x}{\partial _x} + {u_y}{\partial _y}} \right){u_z} + \Omega{u_y} = 0. \\
\end{array}
\end{equation}
Note that we have dropped the $(i,e)$ indices for simplicity and have used the following scalings for the normalization
\begin{equation}\label{nm}
\nabla \to \frac{1}{\lambda_i}\bar \nabla,\hspace{2mm}t \to \frac{{\bar t}}{{{\omega _{pi}}}},\hspace{2mm}{n_{(i,e)}} \to {n_0}{\bar n_{(i,e)}}, \hspace{2mm}{\bf{u_{(i,e)}}} \to {c_s}{\bf{\bar u_{(i,e)}}},\hspace{2mm}\phi  \to \frac{E_{Fe}}{e}\bar \phi,
\end{equation}
where, $\omega_{pi}=\sqrt{4\pi e^2 n_0/m_i}$ is the plasma ion frequency, ${\lambda_i} = c_s/\omega_{pi}$ is the ion gyroradius, and $c_s=\sqrt{E_{Fe}/m_i}$ is the quantum ion sound-speed of plasma. Furthermore, we have neglected the term proportional to small fraction $m_e/m_i$ and introduced the electron nonlocality $H = (\hbar \omega_{pi}/E_{Fe})\sqrt{m_i/(2 m_e^*)}$, the normalized Zeeman $\epsilon=\mu_B B_0/E_{Fe}$, and the normalized ion-cyclotron frequency $\Omega=e B_0/(c m_i\omega_{pi})$ parameters.

The reduced equation set may be linearized using a Fourier-transform using a plane wave ${\bf{G}} = {{\bf{G}}_{\bf{0}}} + {{\bf{G}}_{\bf{1}}}{\rm{Exp}}[{\rm{i}}({\bf{k}}\cdot{\bf{r}} - {\rm{\omega t}})]$ with ${\bf{G}} \equiv {\rm{[n,}}{{\rm{u}}_{\rm{x}}}{\rm{,}}{{\rm{u}}_{\rm{y}}}{\rm{,}}{{\rm{u}}_{\rm{z}}}{\rm{]}}$, ${{\bf{G}}_{\bf{0}}} \equiv {\rm{[1,0,0,0]}}$, and ${\bf{k}} \equiv [k_{\parallel},k_{\perp}]$ ($k_{\parallel}^2+k_{\perp}^2=1$) to give the first-order perturbation relations, as
\begin{equation}\label{red}
\begin{array}{l}
- \omega {n_1} + {k_\parallel }u_{x1} + {k_ \bot }u_{y1} = 0, \\
- \omega u_{x1} =  - \frac{{{H^2}}}{2}{k_\parallel }{n_1} - \frac{{2}}{3}{k_ \parallel }{n_1}, \\
- \omega u_{y1} =  - \frac{{{H^2}}}{2}{k_ \bot }{n_1} - \frac{{2}}{3}{k_ \bot }{n_1} + \frac{{3{\epsilon^2}}}{2}(1 - \frac{{{\mu ^2}}}{3}){k_ \bot }{n_1} + \frac{\Omega }{{\bf{i}}}u_{z1}, \\  - \omega u_{z1} =  - \frac{\Omega}{\bf{i}}u_{y1}. \\
\end{array}
\end{equation}
Consequently, the dispersion relation, obtained from the above linear equation-set can be calculated as
\begin{equation}\label{dis1}
\begin{array}{l}
k_\parallel ^2(4 + 3{H^2})({\omega ^2} - {\Omega ^2}) + {\omega ^2}\left[ {k_ \bot ^2(3{H^2} + 3{\epsilon^2}({\mu ^2} - 3) + 4) - 6({\omega ^2} - {\Omega ^2})} \right] = 0,
\end{array}
\end{equation}
with the two distinct upper (fast) and lower (slow) modes, $\omega_{\pm}$ respectively, given as
\begin{equation}\label{dis2}
\begin{array}{l}
\omega _\pm ^2 = \frac{1}{12}\left[{\Delta  \pm \sqrt {{\Delta ^2} - 24{k_{\parallel}^2}(4 + 3{H^2}){\Omega ^2}} }\right],\hspace{3mm}\Delta  = (4 + 3{H^2}) - 3{k_{\perp}^2}{\epsilon^2}(3 - {\mu ^2}) + 6{\Omega ^2}.
\end{array}
\end{equation}
It is evident that for $\Delta^2<24{k_{\parallel}^2}(4 + 3{H^2}){\Omega ^2}$ the two modes become unstable in Eq. (\ref{dis2}). This condition sets a criteria for the stability of plasma waves based on four independent parameters, namely, the equilibrium number-density, $n_0$, the magnetic-field strength, $B_0$, the fractional effective-mass parameter, $\mu$, and the angle of wave propagation with respect to the magnetic-field, $\theta=\arccos k_{\parallel}=\arcsin k_{\perp}$. The plasma fractional parameters may be related to these parameters as; $ \epsilon \simeq 1.587 \times {10^6}B_0{n_0^{ - 2/3}}\mu^{-1}$, $H \simeq 7200.5{n_0^{ - 1/6}}\mu^{-1/2}$, and $\Omega \simeq 7.277B_0n_0^{-1/2}$. In the proceeding section we will evaluate the stability of waves based on these independent parameters in different quantum plasma circumstances, such as magnetized laboratory and astrophysical degenerate plasmas. \textbf{It is tempting to find the classical dispersion relation from Eq. \ref{dis1} by letting $\hbar\rightarrow 0$ (or equivalently by letting $H\rightarrow 0$). However, as has been pointed out by Haas et al. \cite{haas2} one can not obtain the classical limit of the dispersion relation in this way, since in the classical limit, the equation of state of the quantum plasma which has been used in this model also vanishes through the Fermi-energy which is proportional to the plank constant.}

\section{Polar Mode-Dispersion and Heavy-Fermion Instability}\label{extreme}

Figure 1 depicts the low-field polar dispersion curves for laboratory degenerate plasmas, i.e., ($n_0\simeq 10^{21}/cm^3$ and $B_0\simeq 30T$). Figures 1(a) and 1(b) compares the variation of dispersion for lower and upper modes of QHD waves due to change in the strength of ambient magnetic field. It is observed that, the lower mode frequency increases due to increase in the magnetic field strength, particularly, in direction perpendicular to the magnetic field direction, while, the upper mode frequency it stays unchanged. Figures 1(c) and 1(d) remarked that, as the fractional effective-mass parameter increases the lower mode frequency does not change for $\mu<\mu_{cr}=\sqrt{3}$, while, it decreases for large increase in the value of this parameter. It is noted that, the fields of order $\simeq 40T$ may be easily accessible in high-field laboratories. It is further revealed that both lower and upper mode frequencies are always stable in low-field degenerate environments.

Figures 2 and 3 shows environment of astrophysical degenerate ($n_0\simeq 10^{22}/cm^3$) under medium-field strength ($B_0\simeq 10^4$) which may be present in magnetosphere of a highly magnetized star such as pulsar or magnetar \cite{bing}. It is observed from Figs. 2 and 3 that in medium-field limit the situation is quite different from that in low-field limit shown in Fig. 1. The increase in the magnetic field strength in this case affects both lower (Fig. 2(a)) and upper (Fig. 3(a)) frequency modes. In fact the increase of the magnetic field strength increases the frequency for lower mode, while, it decreases the frequency for the upper one, leading to instability in direction perpendicular to the field for some higher field strength in both frequency modes. From Figs. 2(b) and 3(b) it is further remarked that the increase in the fermion density has quite opposite effects on lower/upper modes compared to the variation in magnetic field strength shown in Figs. 2(a) and 3(a).

On the other hand, Figs. 2(c) and 2(d) remark distinctive behavior of polar dispersion of lower mode in intermediate-field regime. It is observed that, for $\mu>\mu_{cr}$ (Fig. 2(c)) the mode instability unlike for $\mu<\mu_{cr}$ does not occur. The instability of this kind occurs for very small values of the effective-mass parameter and is due to presence of heavy-fermions in the degenerate magnetized plasma. However, both effective-mass parameter region confirm that, the frequency of this mode decreases as the value of $\mu$ departs from the critical value, $\mu_{cr}=\sqrt3$. Moreover, Figs. 3(c) and 3(d) indicate that the upper mode follows similar feature in opposite direction, i.e., distinct differences in effective-mass regimes follow but the frequency in this mode increases as the value of $\mu$ departs from that of the critical. However, the instability is about to set-in for heavy-fermion $\mu$-values.

Figure 4 depicts the instability growth-rate in polar view for the medium-field regime for lower and upper frequency modes corresponding to the heavy-fermion limit. For the lower mode (Fig. 4(a)) it is remarked that, the increase in the value of fractional effective-mass parameter decreases the instability growth-rate in direction perpendicular to the external magnetic field. However, for the case of the upper mode the instability growth is inclined towards the perpendicular direction to the field decreasing slightly in magnitude as the effective mass parameter value increases in the heavy-fermion regime.

In fig. 5 we have evaluated the extreme regime for high electron number-density ($n_0\simeq 10^{28}/cm^3$) under high magnetic-field ($B_0\simeq 10^{8}T$). This situation may be encountered for a strongly magnetic neutron-star (NS) crust. Figures 5(a) and 5(b) indicate the similar features as Figs. 2(a) and 3(a) for the intermediate field strength but highly unstable lower and upper modes in the high field limit. The frequency change in this regime, however, is observed to be mainly confined to the magnetic poles. The polar heavy-fermion instability growth-rates for the two mode in this case are also resemble that of intermediate-field shown in Figs. 4 but somehow pronounced. It is concluded, generally, that critically high-fields present in dense NS crusts cause the magnetosonic waves to be confined in the direction of the ambient strong field, namely, the magnetic poles.

\textbf{It is evident that, the instability presented in this work is due to the interplay between electron spin and orbital magnetization. In other words, the Pauli spin magnetization tends to destabilize the plasma, whereas, the Landau orbital magnetization has the stabilization effect. Therefore, the instability arises when the Landau orbital magnetization or the plasma diamagnetism becomes vanishingly small compared to the Pauli paramagnetism.}

\section{Summary and Conclusions}\label{con}

We studied the propagation the linear propagation of normal modes of QHD waves in a double-degenerate plasma using standard linearizing method. Three different regimes, namely, low intermediate and magnetic field strengths was considered and found that in the laboratory-scale field regime no instability occurs unlike those of intermediate and high fields. We have found a new instability affecting both lower and upper mode frequencies appearing in the heavy-fermion ($\mu<\sqrt{3}$) limit so that the responses of the two modes to fractional effect-mass change in this regime is quite opposite to each other. Furthermore, it was shown that, the heavy-fermion instability due to extremely high magnetic field such as that present in a neutron-star crust can lead to propagations confines to the magnetic poles. This study can be appropriate for investigating wave dynamics in the linear limit for both laboratory and astrophysical environments possessing high magnetic fields.

\newpage

\textbf{FIGURE CAPTIONS}

\bigskip

Figure-1

\bigskip

(Color online) Variation in polar dispersions for lower- and upper frequency modes for the laboratory scale low density and low magnetic field-strength regime with respect to change in various plasma parameters, namely, the magnetic field strength,$B$, and the fractional effective-mass parameter, $\mu$, with variation in one plasma parameter at each plot. The thicknesses of dispersion-curves vary according to change in the varied plasma parameter.

\bigskip

Figure-2

\bigskip

(Color online) Variation in polar dispersions for lower-frequency mode for the intermediate scale magnetic field-strength and electron number-density regime with respect to change in various plasma parameters, namely, the electron number-density, $n_0$, fractional effective-mass parameter, $\mu$, and the magnetic field strength,$B$, with variation in one plasma parameter at each plot. The thicknesses of dispersion-curves vary according to change in the varied plasma parameter.

\bigskip

Figure-3

\bigskip

(Color online) Variation in polar dispersions for upper-frequency mode for the intermediate scale magnetic field-strength and electron number-density regime with respect to change in various plasma parameters, namely, the electron number-density, $n_0$, fractional effective-mass parameter, $\mu$, and the magnetic field strength,$B$, with variation in one plasma parameter at each plot. The thicknesses of dispersion-curves vary according to change in the varied plasma parameter.

\bigskip

Figure-4

\bigskip

(Color online) Variation in polar instability growth-rate of lower- and upper-frequency mode for a intermediate scale magnetic field-strength and electron number-density regime with respect to change in the fractional effective-mass parameter, $\mu$. The thicknesses of dispersion-curves vary according to change in the varied plasma parameter.

\bigskip

Figure-5

\bigskip

(Color online) Variation in polar dispersions and polar instability growth-rate for lower- and higher frequency modes for neutron-star crust scale magnetic field-strength and electron number-density regime with respect to change in various plasma parameters, namely, the electron number-density, $n_0$, fractional effective-mass parameter, $\mu$, and the magnetic field strength,$B$, with variation in one plasma parameter at each plot. The thicknesses of dispersion-curves vary according to change in the varied plasma parameter.

\end{document}